 \documentclass[a4paper]{aa}

\usepackage{graphicx,times}

\def\''{\lq\lq}

\def\gsim{\lower.4ex\hbox{$\;\buildrel >\over{\scriptstyle\sim}\;$}}
\def\lsim{\lower.4ex\hbox{$\;\buildrel <\over{\scriptstyle\sim}\;$}}
\def\~  {$\sim$}

\def\newpage{\vfill\eject}

\def\apj{{ApJ}}
\def\apjs{{ Ap. J. Suppl.}}
\def\apjl{{ Ap. J. Letters}}

\def\aa{{A\&A}}

\def\an{{Astron. Nachr.}}

\def\bib{\item{}}
\renewcommand{\vec}[1]{\mbox{\boldmath$#1$}}

\def\L{$\Lambda\;$}
\def\L{$\Lambda$}

\input epsf

\begin{document}


\title{Differential rotation of  main sequence F stars}
 \author{M. K\"uker and G. R\"udiger}
\offprints{M.~K\"uker}
\institute{Astrophysikalisches Institut Potsdam,  An der Sternwarte 16,
D-14482 Potsdam, Germany}
\date{Received  / Accepted }
%
%
\abstract{
The differential rotation of a 1.2 $M_\odot$ zero age MS star (spectral type F8) is computed  and the results are compared with those from a similar  model of  the Sun. The rotation pattern is determined by solving the Reynolds equation including the convective energy transport. The latter is anisotropic due to the Coriolis force causing a horizontal temperature gradient of ~ 7 K between  the poles and the equator.  Comparison of the  transport mechanisms of angular momentum (the eddy viscosity,  the \L-effect  and the meridional flow) shows that for the F star the \L-effect is the most powerful transporter for rotation periods of 7 d or less. In the limit of very fast rotation the \L-effect is balanced by the meridional flow alone and the rotation is nearly rigid.  The rotation pattern found for the F star is very similar to the solar rotation law, but the horizontal shear is about twice  the solar value. As a function of the rotation period, the total equator-pole difference of the angular velocity  has a (slight)  maximum at a period of 7 d and (slowly) vanishes in both the limiting cases of very fast and very slow rotation.  A comparison of the solar models with those for  the F-type star shows a much stronger dependence of the differential surface rotation on the stellar luminosity rather than on the rotation rate (see Fig.~\ref{fsom}).
}
 \authorrunning{M.~K\"uker \& G.~R\"udiger}
\titlerunning{Differential rotation of MS F stars}
\maketitle

\section{Introduction}
%
The differential rotation of the solar photosphere has been observed for a long time.
Depending on the method used, the rotation period at the equator is found to be 20--30 percent shorter than at the poles. Over the last 15 years helioseismology has revealed that this phenomenon is not restricted to the atmosphere,  granulation, and supergranulation layers but persists throughout the whole convection zone. Moreover, the gas rotation does not show the expected increase with depth. Instead, the pattern observed at the surface is found throughout the whole convection zone while the core appears to rotate rigidly at the same rate as the convection zone does at intermediate latitudes (Thompson et al.~2003).

The solar rotation pattern can be explained with angular momentum transport by
the gas flow in the convection zone. Assuming that Reynolds stress is the only
relevant transporter of angular momentum, K\"uker et al. (1993) found a remarkably good agreement between the results from their model based on the theory of the Reynolds stress and the observed rotation. The model, however, does not explain the  impact of other transporters, especially the meridional flow. Kitchatinov \& R\"udiger (1995) and K\"uker \& Stix (2001) presented a refined model that included both the meridional flow and the convective heat transport. They showed that the flows driven by the differential rotation and a small horizontal temperature gradient due to anisotropic convective heat transport essentially have opposite directions, and roughly balance each other in the solar convection zone.

Hall (1991) found differential rotation for a number of magnetically active stars from the variation of spot rotation periods over the stellar activity cycle.
The variation of stellar rotation periods over the activity cycle was also used to measure the surface differential of active lower main-sequence stars by Donahue et al.~(1996).
 Messina \& Guinan (2003)  derived surface differential rotation from photometric data from a monitoring program that studies a sample of stars that resemble the Sun in an earlier state of its evolution.

The number of single main-sequence stars to which Doppler imaging   has been successfully applied is  still small (Strassmeier 2002).
For sufficiently fast rotators, surface differential rotation can be detected, however,  from the broadening of spectral lines. Reiners \& Schmitt (2003a,b) carried out measurements for F stars with moderate and short rotation periods. They found differential rotation to be much more common for stars with moderate rotation rates than for very rapid rotators.

Here  we  present a model for the differential rotation of a main sequence star of spectral type F8. This star, with 1.2 solar masses, lies just below the maximum mass for main-sequence stars with outer convection zones and therefore represents the upper end of the lower main-sequence in the context of differential rotation and stellar activity. With a convection zone depth of 160,000 km and a surface gravity roughly equal to the solar value the main difference from the Sun is the luminosity, which is 1.7 times the solar value. 
%
\section{The Model}
%
%
The basic setup is the same as in K\"uker \& Stix (2001) and very similar to Kitchatinov \& R\"udiger (1999). The stellar parameters are taken from a model of a 1.2 $M_\odot$ ZAMS star computed by Granzer (2002). The fractional depth of the convection zone is 20 $\%$ of the stellar radius of 1.13 solar radii. While the depth of the convection zone is similar to that of the Sun, the greater luminosity enforces a larger convective heat flux. The density stratification is steeper than in the solar convection zone, leading to smaller values of the density and pressure scale heights, and hence smaller values of the mixing-length. As a consequence, the convection velocity must be larger than in the solar convection zone.

We treat rotation and large-scale meridional flow as a perturbation with the stratification from a standard stellar evolution code as the unperturbed state. For rotation periods longer than one day the deviation from spherical symmetry is small. The unperturbed state is therefore assumed spherically symmetric.

%
The equations to be solved are the conservation laws of angular and linear momentum, and the equation of convective heat transport.
With the mean-field ansatz, $\vec{u}=\vec{\bar{u}}+\vec{u'}$, the equation of motion for the mean velocity field $\vec{\bar{u}}$ reads
\begin{equation} \label{reynolds}
  \rho \left [ \frac{\partial \vec{\bar{u}}}{\partial t}
      + (\vec{\bar{u}}\cdot \nabla)
       \vec{\bar{u}} \right ] = -  \nabla \cdot \rho Q
	   - \nabla P + \rho \vec{g},
\end{equation}
with  the one-point correlation tensor
$
   Q_{ij} =  \langle u_i' u_j' \rangle.
$
We define 
\begin{equation}
  \vec{\beta}=\frac{\vec{g}}{c_p}-\vec{\nabla}T
\end{equation}
and split it into its horizontal average and the deviation from that average:
 \begin{equation}
    \vec{\beta}(r,\theta) = \vec{\beta}^1(r) + \vec{\beta}^2(r,\theta),
  \end{equation}
where
 \begin{equation}
\vec{\beta}^1(r)=\frac{1}{2}\int_0^\pi \vec{\beta} \sin \theta { d}\theta.
   \end{equation}
The convective heat diffusion coefficient is defined in terms of $\vec{\beta}^1$, using the form derived by Kitchatinov \& R\"udiger (1999) instead of standard mixing-length theory:
\begin{equation} \label{chi_iso}
 \chi_{\rm t} =  \frac{\tau_{\rm corr} g \alpha_{\rm MLT}^2 H_p^2}{12 T} \beta^1_r
\end{equation}
A similar expression holds for the turbulence viscosity,
\begin{equation} \label{nu_iso}
 \nu_{\rm t} =  \frac{\tau_{\rm corr} g \alpha_{\rm MLT}^2 H_p^2}{15 T}\beta^1_r
\end{equation}

The heat transport is then described by the transport equation
\begin{equation} \label{heat}
\rho T \frac{\partial s}{\partial t}
          = - \nabla \cdot  (\vec{F}^{\rm conv} + \vec{F}^{\rm rad}) +\rho c_p \vec{u} \cdot \vec{\beta}^2,
\end{equation}
with the convective heat flux,
\begin{equation} \label{convheat}
 F_{i}^{\rm conv} = -\rho c_p \chi_{ij}\beta_j.
\end{equation}
 The radiative heat flux reads
\begin{equation}
 F_i^{\rm rad} = - \frac{16 \sigma T^3}{3 \kappa \rho} \nabla_i T,
\end{equation}
with the opacity $\kappa$. The advection term in Eq.~(\ref{heat}) contains  $\vec{\beta}^2$ instead of $\vec{\beta}$ to remove the convective instability from the system. The radiative and convective transport terms account for the entire radial heat transport. Heat advection by the large-scale meridional flow is included only to allow for a back-reaction against the baroclinic term, $\frac{1}{\rho^2}(\nabla \rho \times \nabla p)_\phi $  in Eq.~(\ref{reynolds}). Assuming hydrostatic equilibrium,
\begin{equation}
  \nabla P = - \vec{g} \rho,
\end{equation}
we can rewrite the baroclinic term in terms of the specific entropy:
\begin{equation}
  \frac{1}{\rho^2}(\nabla \rho \times \nabla P)_\phi =
 - \frac{1}{c_p \rho}(\nabla s \times \nabla P)_\phi \approx
 - \frac{g}{r c_p} \frac{\partial s}{\partial \theta}.
 \end{equation}
The tensor $\chi_{ij}$  contains the anisotropy of the convective heat transport  (Kitchatinov et al.~1994; K\"apyl\"a, Korpi \& Tuominen  2004,  R\"udiger et al. 2004).
%
%
%
\section{Results}
The model  has first been applied to the Sun by K\"uker \& Stix (2001). With the original value of 4/15 for $c_\nu$ and a value of 5/3 for the mixing-length parameter, a value of 0.2 was obtained for the normalised horizontal shear,
\begin{equation}
\frac{\delta \Omega}{\Omega} = \frac{\Omega_{\rm eq}-\Omega_{\rm pole}}{\bar{\Omega}_{eq}},
\end{equation}
where $\Omega_{\rm eq}$ and $\Omega_{\rm pole}$ are the rotation rates at the equator and the poles. As the resulting horizontal shear is  smaller than the observed 30 $\%$, we treat the viscosity parameter $c_\nu$ as a free parameter and vary it to improve the reproduction of the observed rotation law. The turbulent heat transport coefficient $c_\chi$ then results from the requirement that the Prandtl number keeps its value of 0.8. Test computations have shown that for the Sun  the latitudinal shear is 20 $\%$  for $c_\nu = 4/15$  and 30 $\%$  for  $c_\nu=0.15$. We therefore carry out the computations for our model F star with this value.
%
\subsection{The rotation law}
\begin{figure*}[ht]
\mbox{
\includegraphics[width=4.0cm]{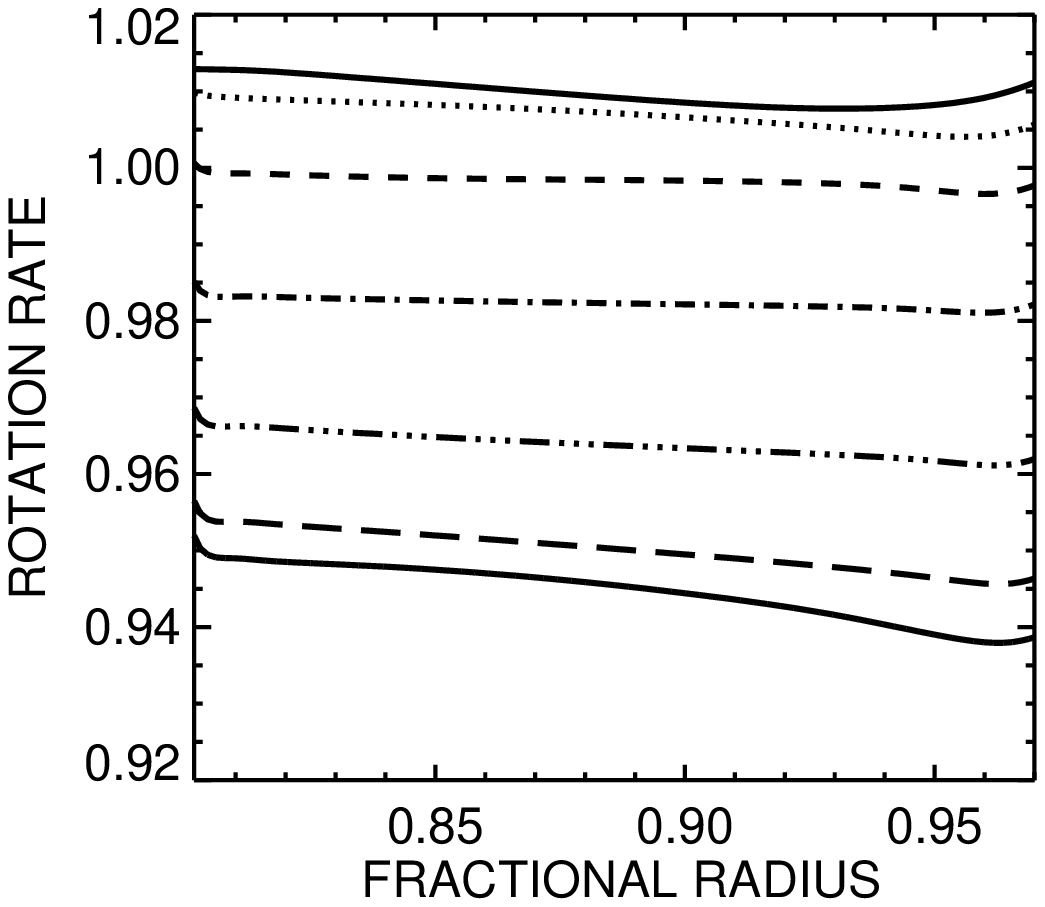}
\includegraphics[width=4.0cm]{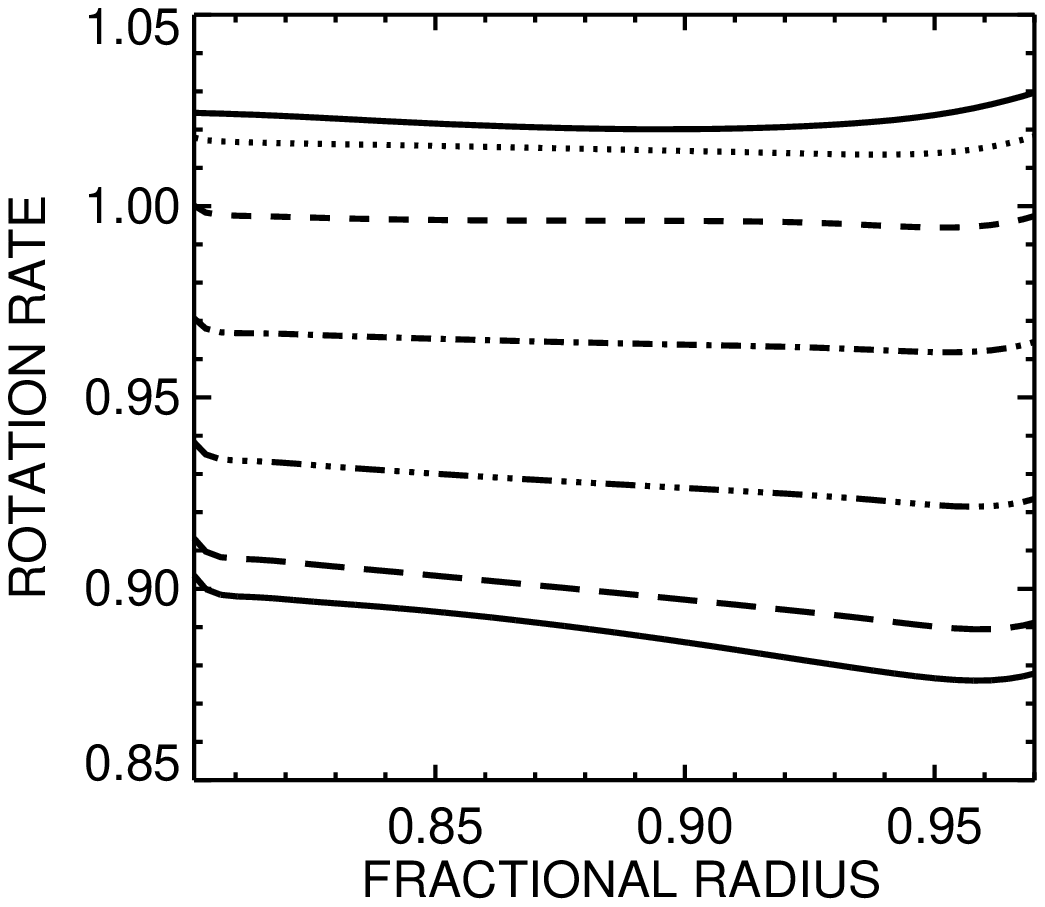}
\includegraphics[width=4.0cm]{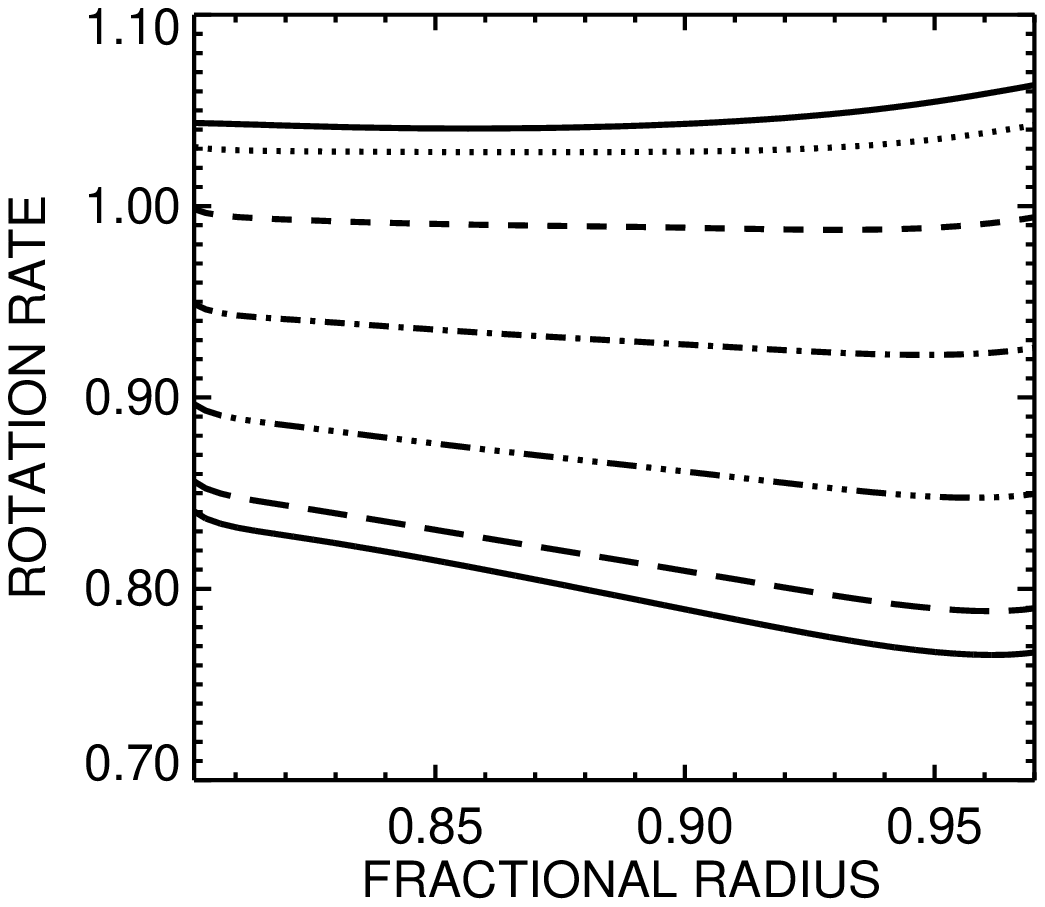}
\includegraphics[width=4.0cm]{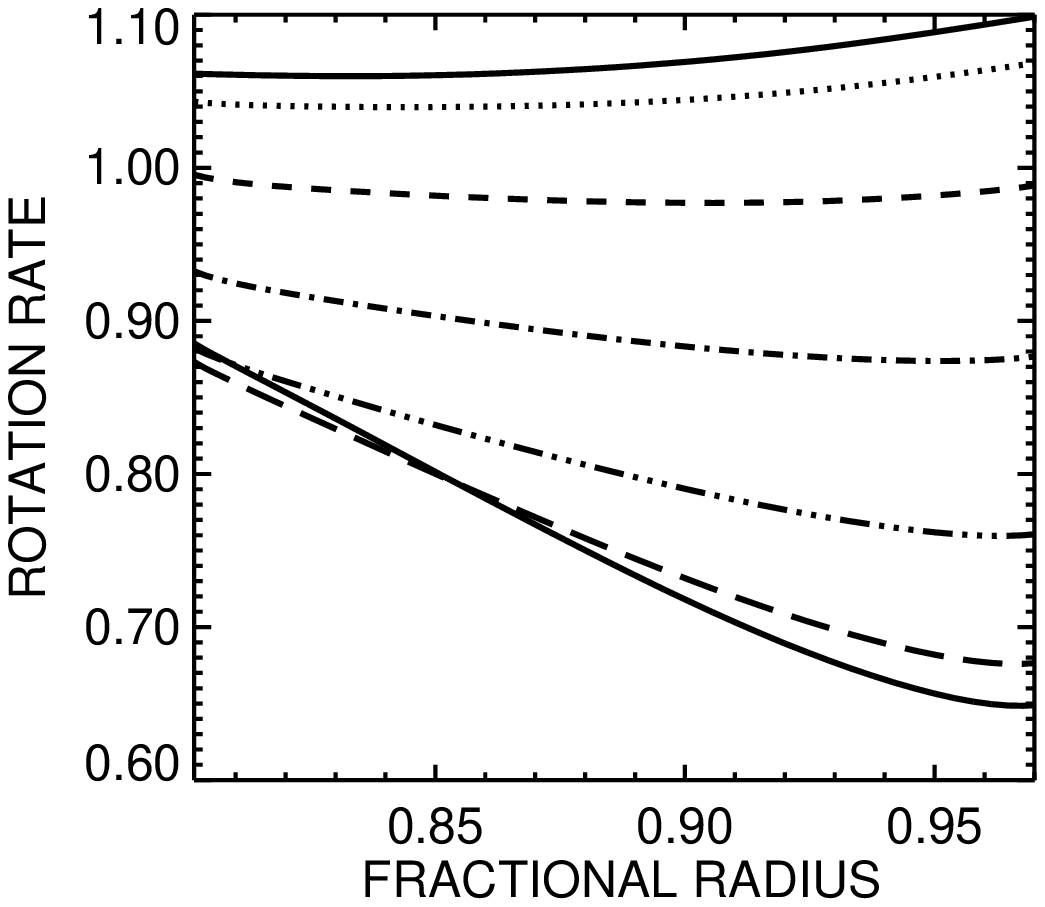}
}
\caption{
Normalised rotation rate as a function of the fractional stellar radius for the 0$^\circ$, 15, 30, 45, 60, 75, and 90$^\circ$  latitude (from top to bottom in each diagram) for rotation periods of 4 d, 7 d, 14 d, and 27 d (from left to right).
}
\label{f4}
\end{figure*}

Figure \ref{f4} shows the rotation  pattern for the F star model with $c_\nu = 0.15$ and $\alpha_{\rm MLT}=5/3$ and
rotation periods of 4 d, 7 d, 14 d, and 27 d. We define the mean angular velocity as the ratio of the angular momentum and the moment of inertia, i.e.
\begin{equation}
    \label{shear1}
 \bar{\Omega} = \frac{\int \rho r^4 \sin^3 \theta \Omega { d}r { d}\theta}{\int \rho r^4 \sin^3 \theta { d}r { d}\theta}.
 \end{equation}
The mean rotation period is then
\begin{equation}
 P_{\rm rot} =  \frac{2 \pi }{ \bar{\Omega}},
\end{equation}
which (naturally) corresponds to the rotation period at a latitude of 30$^\circ$  for a solar-type rotation law.

\begin{figure*}[ht]
\mbox{
\includegraphics[width=4.0cm,height=6cm]{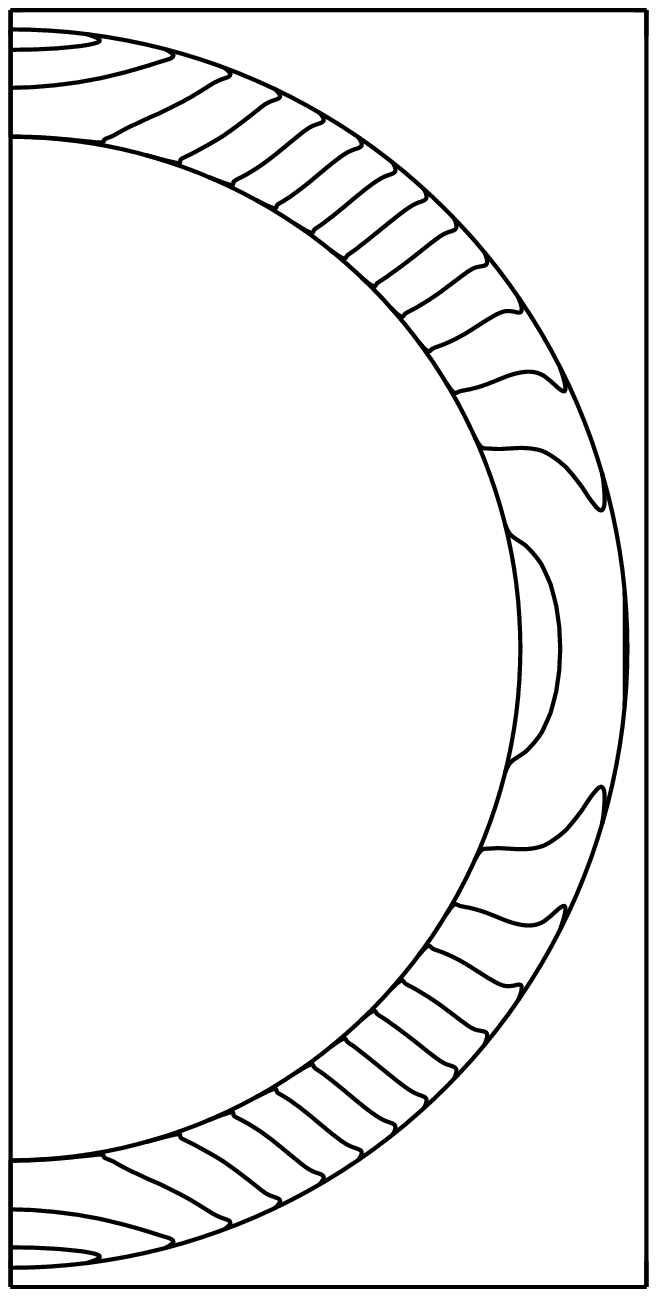}
\includegraphics[width=4.0cm,height=6cm]{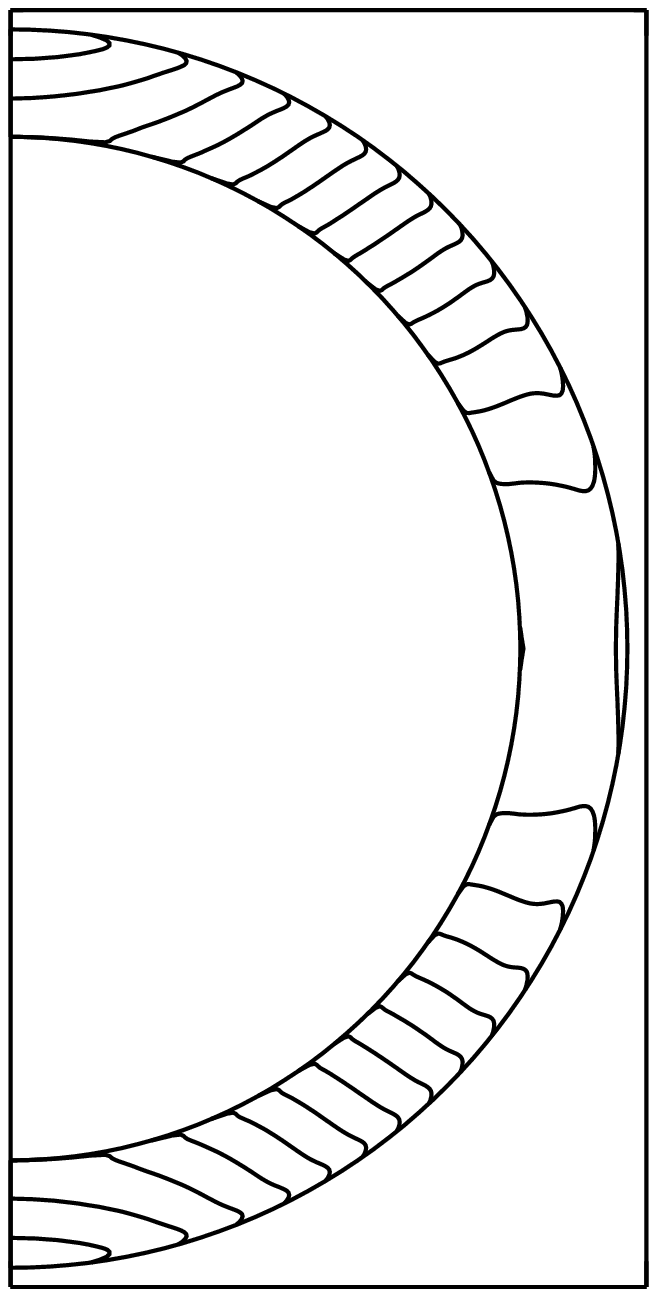}
\includegraphics[width=4.0cm,height=6cm]{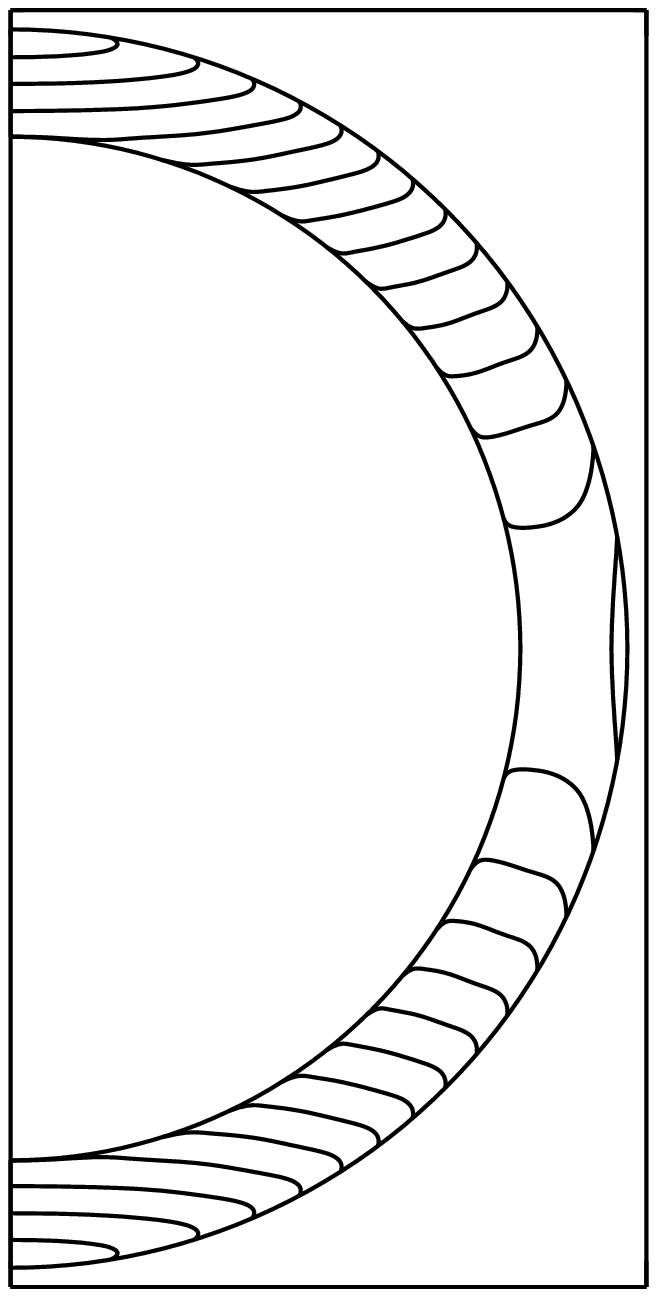}
\includegraphics[width=4.0cm,height=6cm]{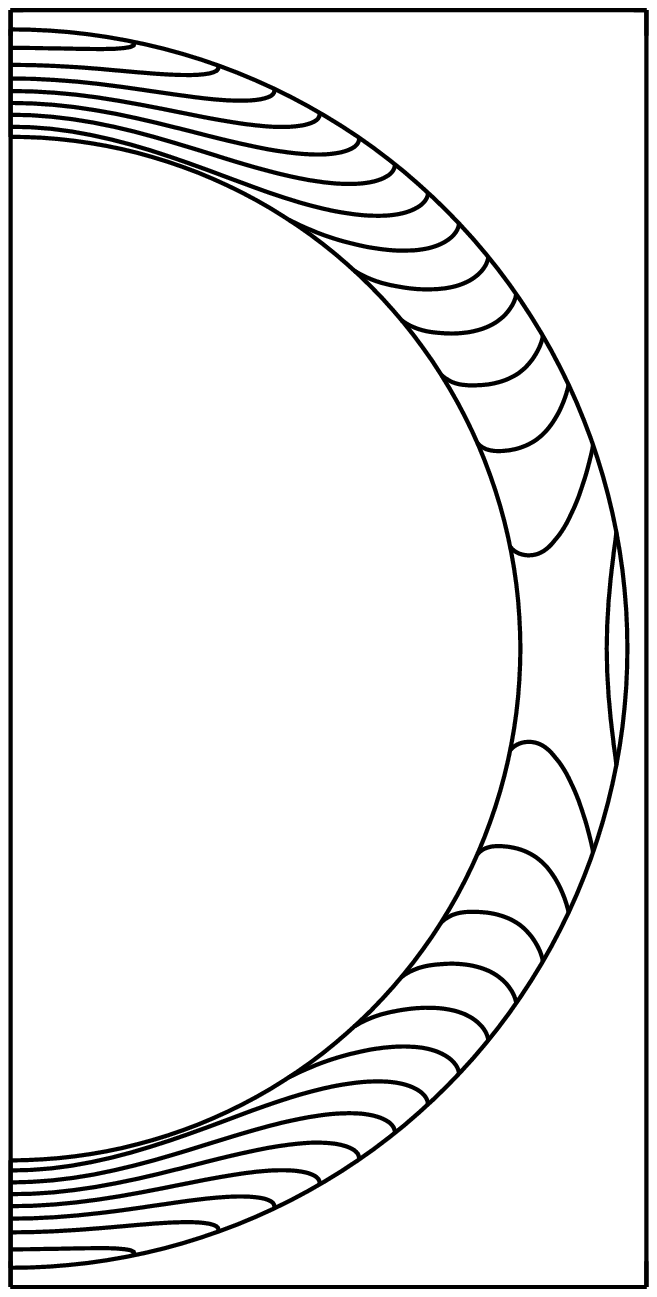}
}
\caption{
Isocontours of the rotation rate for average rotation periods of 4 d, 8 d, 16 d, and 27 d (from left to right).
}
\label{f1}
\end{figure*}

In Fig.~\ref{f4} the rotation rate is plotted vs.~radius for latitudes at 0$^\circ$  (equator), 15, 30, 45, 60, 75, and 90$^\circ$  (poles). In all cases the equator rotates faster than the poles (solar-type differential rotation), but the amplitude of the relative shear varies  with the rotation rate. The model with the fastest rotation  yields the most rigid rotation law. Note also that for fast rotation all lines are practically horizontal, i.e.~there is no radial shear, while the model with a rotation period of 27 d shows a substantial decrease of the rotation rate with increasing radius at high latitudes. This change of the rotation pattern can also be seen in the contour plots shown in Fig.~\ref{f1}. While the isocontours are mainly radial for fast rotation the slow rotation case (27 d) shows a disk-shaped pattern at high latitudes.
\begin{figure}
\includegraphics[width=8.0cm]{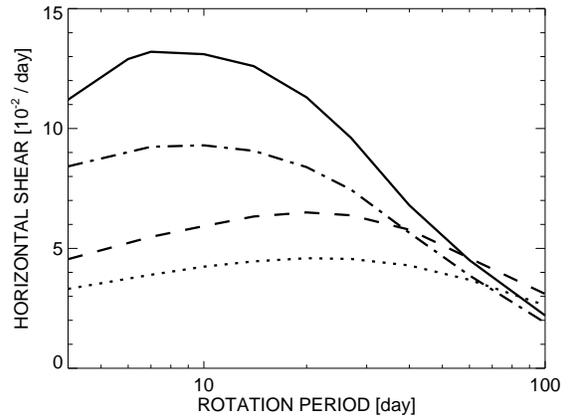}
\caption{The total horizontal shear for the F8 star as a function of the rotation period. Solid line: $c_\nu=0.15,$ dot-dashed line: $c_\nu=4/15.$
Dashed line: solar-type star, $c_\nu=0.15.$ Dotted line: solar-type star, $c_\nu=4/15.$
\label{fsom}
}
\end{figure}
Figure \ref{fsom} shows the total horizontal shear, $\delta \Omega$, as a function of the rotation period for the F8 star and the Sun for both values of the viscosity parameter $c_\nu$. All four curves show a decrease of the shear with increasing rotation period for slow rotation. For rapid rotation, the behaviour is the opposite: The rotation becomes more rigid as the period decreases, i.e.~the rotation gets faster. Between the two limiting cases the total horizontal shear assumes its maximum value. The period at which the maximum is reached is about one month in case of the Sun, and 19 d for the F star. None of the curves shows a sharp peak. There is always a broad range of nearly constant shear around the maximum. For both stars, the shear strongly depends on the viscosity parameter.
\subsection{The meridional flow}
\begin{figure*}[htb]
\mbox{
\includegraphics[width=4.0cm,height=6cm]{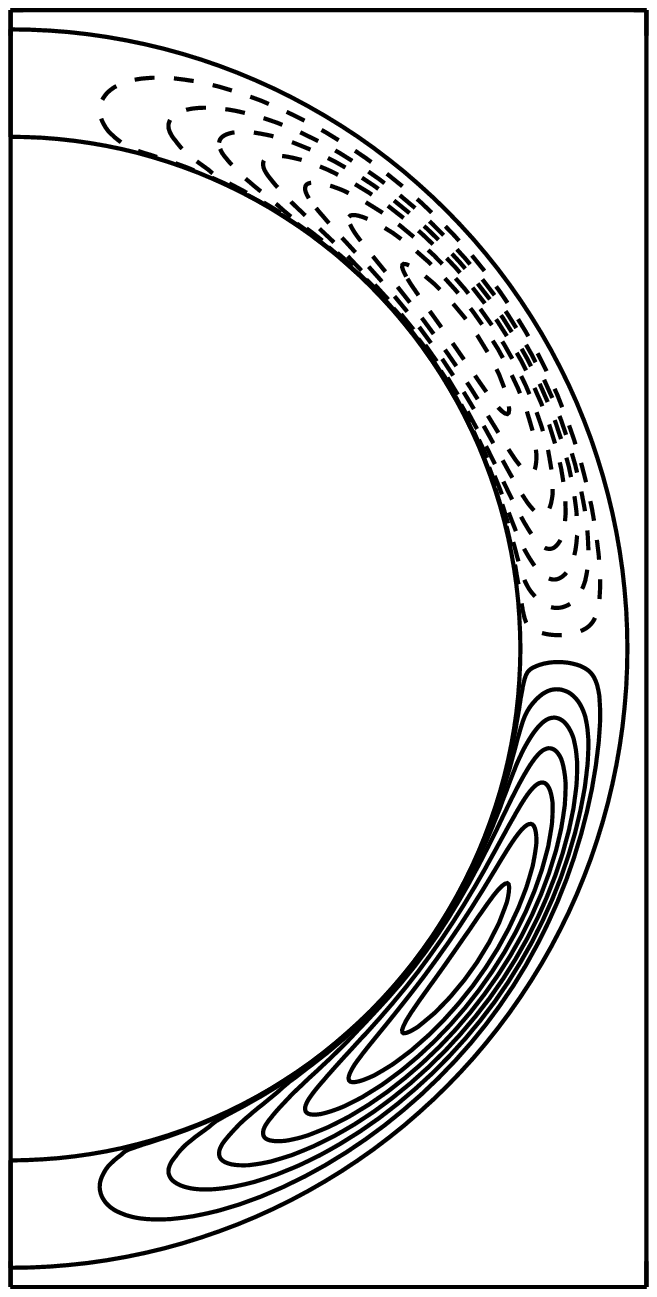}
\includegraphics[width=4.0cm,height=6cm]{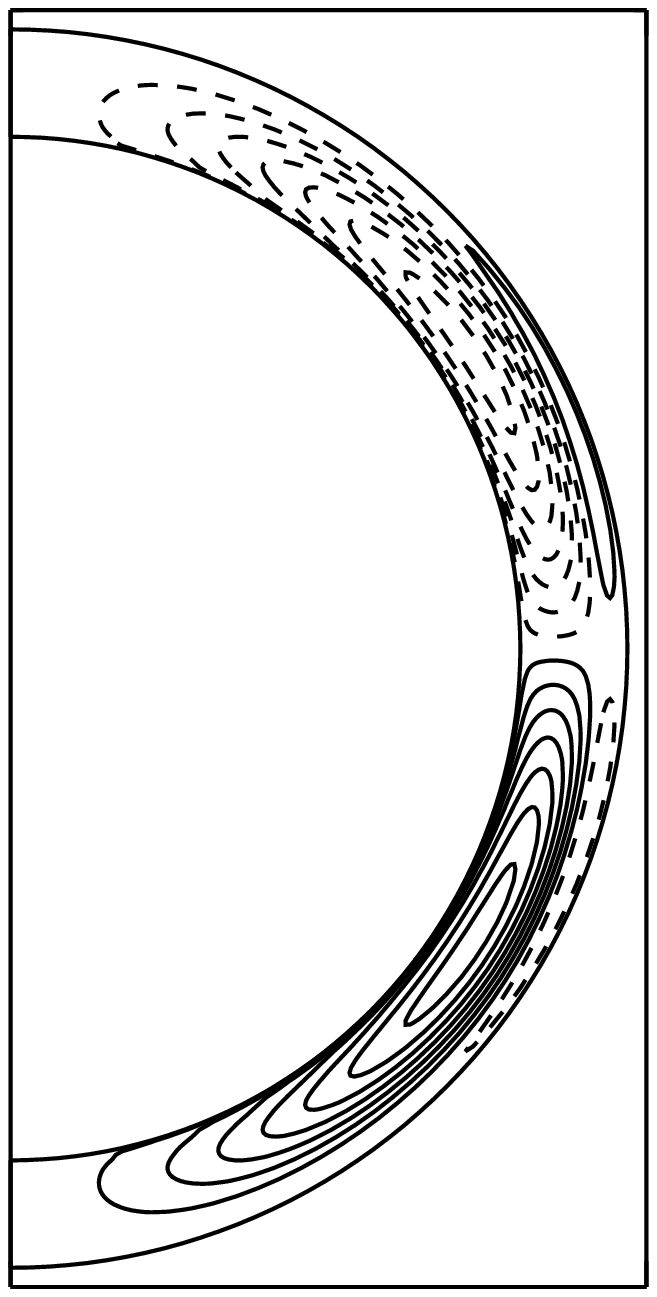}
\includegraphics[width=4.0cm,height=6cm]{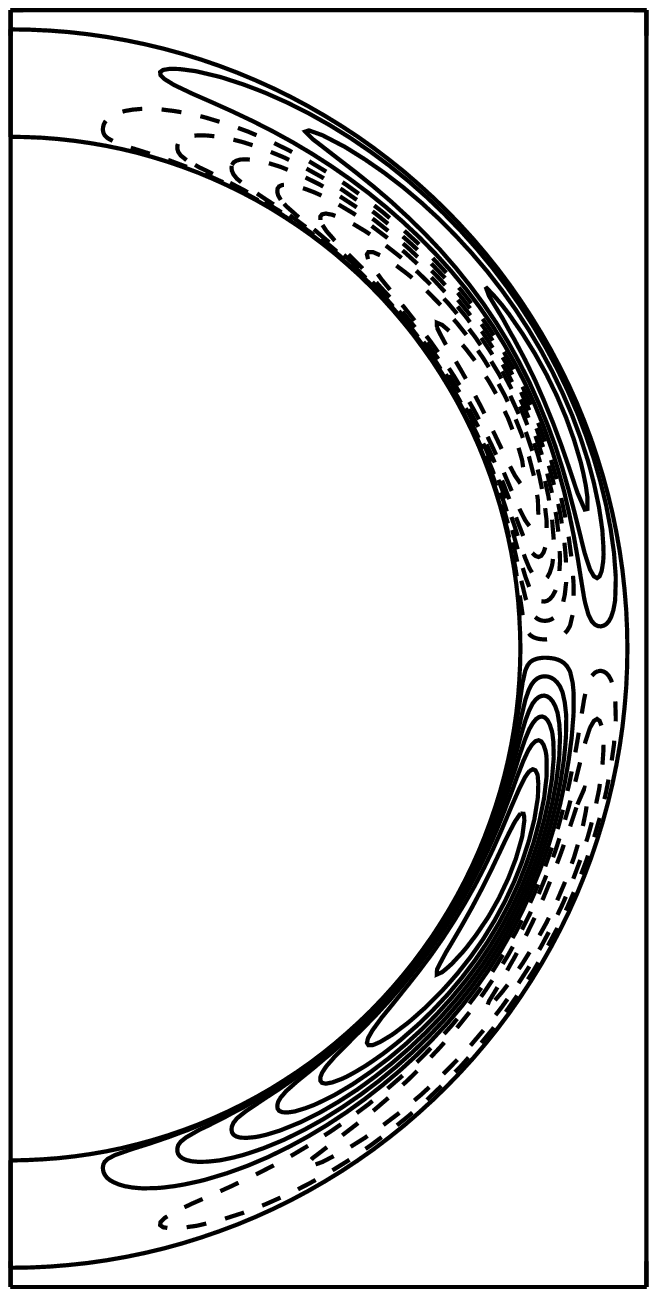}
\includegraphics[width=4.0cm,height=6cm]{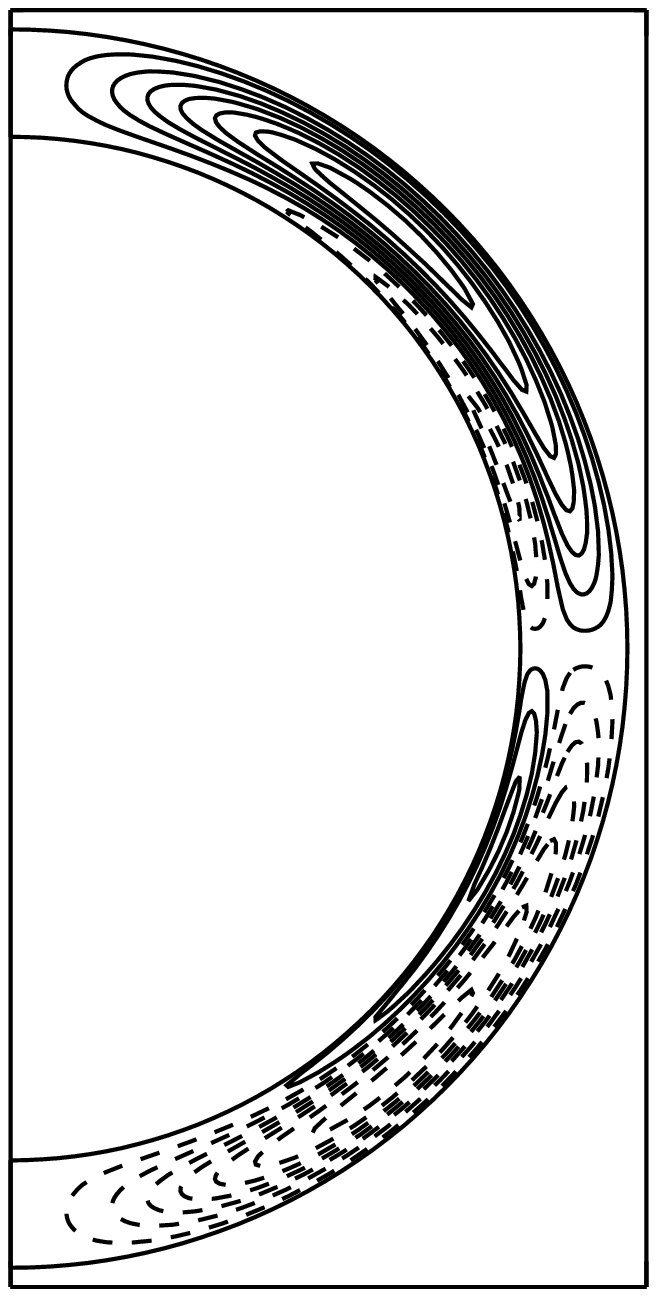}
}
\caption{
Stream lines of the meridional flow for  rotation periods of 4 d, 7 d, 14 d, and 27 d (from  left to right).
}
\label{f2}
\end{figure*}
Figure \ref{f2} shows the meridional flow patterns. For  $P=4$ d  the figure shows only one flow cell per hemisphere, with the flow directed towards the poles in the upper part of the convection zone and towards the equator at the bottom.
For a rotation period of 7 d, a second flow cell is  visible. It is directed towards the equator both at the top and the bottom of the convection zone, but towards the poles in the bulk of the convection zone. At a rotation rate of 14 d the upper cell has become larger, and for $P=27$ d it dominates the flow pattern, with the counter-clockwise flow restricted to a thin layer at the bottom.

\begin{figure}[htb]
\includegraphics[width=8.0cm]{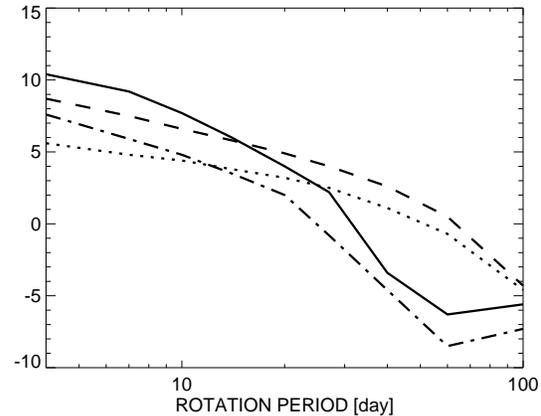}
\caption{Meridional flow speed in m/s at the bottom of the convection zone as a function of the rotation period. Positive value correspond to gas motion toward the equator. Solid line:F8 star, $c_\nu=0.15$, dashed line: F8 star, $c_\nu=4/15.$
Dashed line: solar-type star, $c_\nu=0.15.$ Dotted line: solar-type star, $c_\nu=4/15.$
\label{fsutet}
}
\end{figure}

In Fig.~\ref{fsutet} the maximum flow speed at the bottom of the convection zone is shown. The sign refers to the direction of the flow at the point where it is fastest. A positive sign means that the flow is toward the equator, negative values indicate poleward flow.
For the F8 star and $c_\nu=0.15$ the value of the speed decreases from 10.4 m/s for $P_{\rm rot}=4$ d to --6.3 m/s at $P= 60$ d. The change of the flow direction at the bottom of the convection zone should have dramatic consequences for the stellar dynamo, if indeed the form of the butterfly diagram is dominated by the meridional flow at the bottom of the convection zone. 

 For rotation periods longer than $P_{\rm rot}=60$ d the speed increases as the absolute value decreases. For fast rotation the flow is faster in the $c_\nu=0.15$ case than in the $c_\nu=4/15$ case, while for slow rotation the latter gives the faster flow. 
%
\subsection{Differential temperature}
\begin{figure}[htb]
 \includegraphics[width=4.3cm]{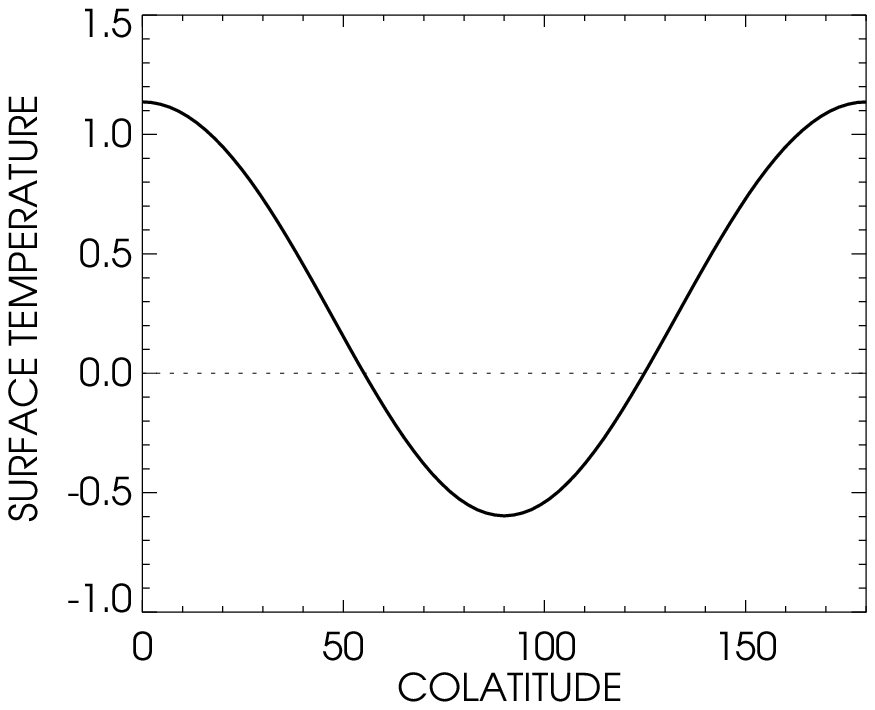}
 \includegraphics[width=4.3cm]{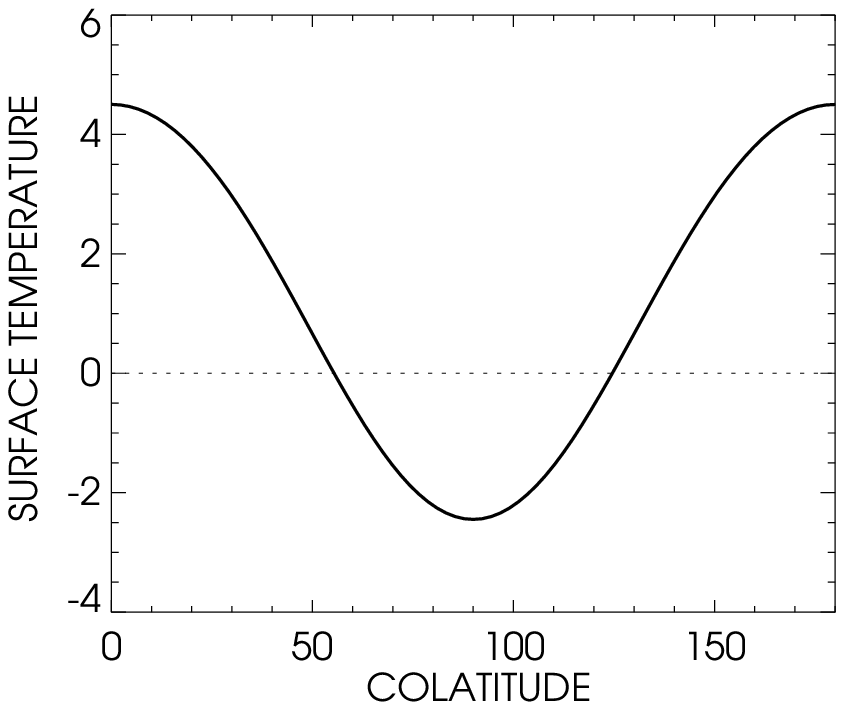}
 \caption{
  \label{tsurf}
  The temperature variation (in K) at the upper boundary. Left: solar-type star, $P_{\rm rot}=27$ d, right: F8 star, $P=10$ d. The horizontal average has been subtracted.
  }
\end{figure}

Figure \ref{tsurf} shows the variation of the temperature with latitude at the upper boundary for the solar-type star and the F8 star. The horizontal average has been subtracted. In both cases the polar region is slightly hotter than the equator. For the G2 star the difference is 1.7 K, for its F8 counterpart 7.0 K. The average temperatures are {$2.8 \times 10^5$ K} and {$1.57 \times 10^5$ K}, respectively.

\section{Discussion}
%
%
For the same set of parameters, the F star shows a  {\em larger difference} between the equatorial and polar rotation rates. As the convection zones are of roughly the same depth, this suggests that differential rotation is determined by the stellar luminosity rather than the rotation rate.
Both mixing length and convective turnover time in the solar convection zone are larger than the corresponding values in the F star convection zone. The opposite holds for the convection velocity. The mixing length in the F star is only slightly smaller than in the solar convection zone, by about as much as the convection zone is shallower. The convection velocity is twice that of the Sun because of the higher luminosity.

A comparison between the two convection zones can be summarised as follows. Since the stellar radii and depths of the convection zone are very similar, the mixing lengths are roughly the same, too. The higher luminosity of the F star requires a larger convective heat flux, therefore larger velocities, and hence shorter convective turnover times. As a result, the difference in the Coriolis number between an F8 star with a 7 d rotation period and a G2 star rotating with a period of 27 d is less than a factor of two, half the ratio of the rotation periods.

\begin{figure}
  \begin{center}
  \includegraphics[width=7.0cm]{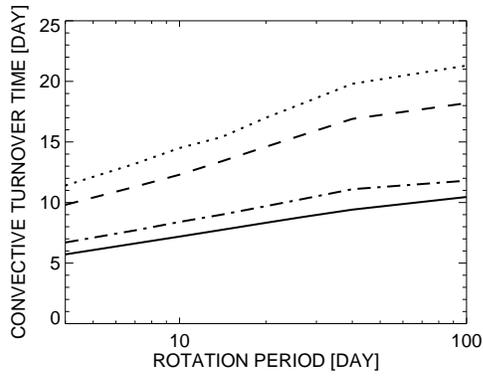}
  \end{center}
 \caption{
 The global convective turnover time as a function of the rotation periods for the F star and the solar-type star.
Solid line: F star, $c_\nu=0.15.$ Dash-dotted line: F star, $c_\nu=4/15.$  Dashed line: solar-type star, $c_\nu=0.15.$ Dotted line: solar-type star, $c_\nu=4/15.$
   \label{ftau}
 }
\end{figure}

Kim \& Demarque (1996) define the global turnover time,
\begin{equation}
  \tau_g=\int_{R_b}^{R_\star} \frac {{ d}r}{v_c},
\end{equation}
as an integral over the entire extent of the convection zone. With a value of 1.86 for the mixing length parameter they find $\tau_c=13.95$ d for a 1.2 $M_\odot$ star.
Figure \ref{ftau} shows this quantity as resulting from our model for the F8 star as a function of the rotation period. While not constant, $\tau_g$ is a slowly varying function of the rotation period. Rapid rotation enforces shorter time scales because the efficiency of the convective heat transport is reduced. As both the convective heat flux and the mixing length are prescribed by the boundary conditions and the stellar stratification, respectively, the only way for the system to adjust  is  by increasing the convection velocity, and hence shortening the turnover time. The values found are smaller than the 13.95 d of Kim \& Demarque (1996), but the curve in Fig.~\ref{ftau} shows an increase towards slower rotation. Using standard mixing-length theory, their model corresponds to ours in the limit of slow rotation. Note that the global convective turnover time is about twice as long as its locally defined counterpart.

%
%
Our model predicts the surface rotation pattern of lower main-sequence stars to depend on luminosity much more than on the rotation rate.
This finding is in agreement with the observed surface differential rotation of AB Dor (Donati \& Collier Cameron 1997), PZ Tel  (Barnes et al.~2000), and LQ Lup (Donati et al.~2000).  These stars, though  rotating much faster than the sun, show surface differential rotation similar to that of the Sun. LQ Lup (RXJ1508.6-4423) is a post T Tauri star of spectral type G2. With a rotation period as short as 0.31 d the surface shear $\delta \Omega = 0.13$ d$^{-1}$ is only about twice the solar $\delta \Omega = 0.07$ d$^{-1}$. AB Dor and PZ Tel are PMS stars of spectral type K0 and K9/G0, respectively, with rotation periods of 0.5 d and 1 d. With $\delta \Omega=0.056$ d$^{-1}$ (AB Dor ) and $\delta \Omega= 0.075$ d$^{-1}$ (PZ Tel) the surface shear values of these stars lie close to the solar value, with the more luminous AB Dor also showing more surface shear.

\begin{figure}
  \includegraphics[width=8.0cm]{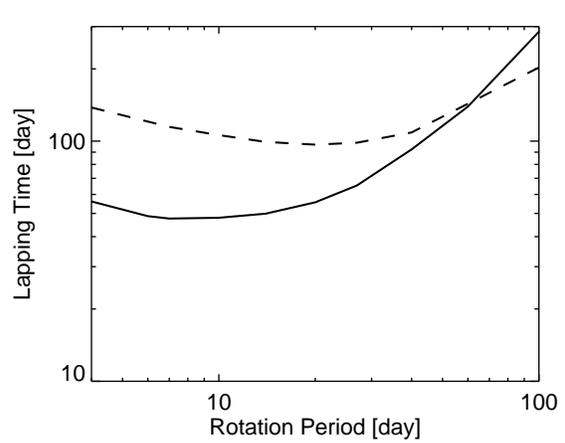}
   \caption{ \label{lapping}
     The lapping time as a function of the rotation period for the F star (solid line) and the solar-type star (dashed). 
   }
\end{figure}

Reiners and Schmitt (2003 a,b) found values between 10 and 30 for the lapping time, $P_{\rm lap}=2\pi/(\delta \Omega)$. Figure \ref{lapping} shows the lapping time vs.~the rotation period for the F and solar-type stars for $c_\nu = 0.15$ from our model. For both stars there is little variation except for very long periods, where the lapping time increases profoundly. The value for the solar-type star is about 100\,d at the period of maximum shear. That of the F star 50\,d, 2--3 times larger than those found by Reiners \& Schmitt for the same interval of rotation periods.

Reiners \& Schmitt found differential rotation to be much more common for F-type stars with moderate rotation rates than for the most rapidly rotating ones. As the method they use detects differential rotation only when the {\em relative } shear, $\delta \Omega / \Omega$ exceeds certain threshold value, their finding  is not surprising. As the {\em total} shear is a slowly varying function of the rotation rate, the relative shear is roughly proportional to the rotation period, i.e.
$
\delta \Omega/\Omega  \propto P_{\rm rot}
$
The non-detection of differential rotation therefore does not rule out its presence for rapid rotators and it is actually predicted by our model.

Several observers have studied the connection between rotation period and surface rotational shear for late-type stars. The result is usually expressed as a power law, i.e.
\begin{equation}
   \delta \Omega \propto P_{\rm rot}^n
\end{equation}
In this representation  a constant shear  reads
$n=0$.
Figure \ref{fsom} shows the surface horizontal shear as a function of the rotation period for the F8 star and for the solar model. The functions $\delta \Omega=\delta \Omega(P_{\rm rot})$ are not power laws, especially not  for the F8 star. A value of 0.5 for the exponent n gives a good fit for the latter with $c_\nu=0.15$ in the interval from 4 d to 7 d while a smaller value of -0.5 is found for periods between 14 d and 20 d. Around $P_{\rm rot} = 10$ d we find $n=0$, which corresponds to the maximum of $\delta \Omega=\delta \Omega(P_{\rm rot})$.

Hall (1991) found $n=-0.21 $ while Donahue et al.~(1996) derived $n=-0.7$ from observations of the Ca II chromospheric lines. Schmitt \& Reiners report a value of $-0.53$, and Messina \& Guinan derive an exponent of $-0.6$ (see R\"udiger \& Hollerbach 2004). The observed values tend to be smaller than the value of 0...0.5  which our model predicts for rapid rotation. A possible explanation is scatter caused by the variety of spectral types in the observed samples.

\begin{figure}
 \begin{center}
  \includegraphics[width=7.0cm]{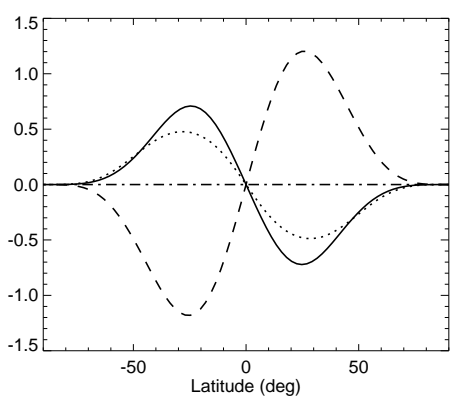}
 \end{center}
  \caption{ Horizontal flux of angular momentum  in the convection zone of the solar model ($P_{\rm rot}=27$d, $c_\nu=0.15$) with the baroclinic term included.
Solid: contribution of the meridional flow. Dashed: \L-effect. Dotted: viscosity. Dash-dotted: total angular momentum flux.
    \label{ftet_sun}
    }
  \end{figure}
\begin{figure*}
 \begin{center}
  \includegraphics[width=7.0cm,height=5cm]{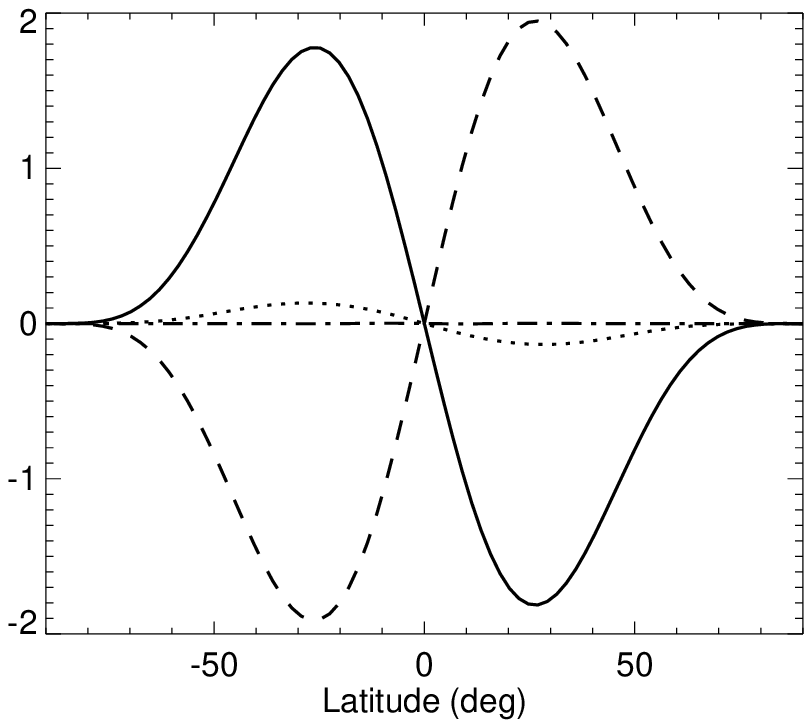}
  \includegraphics[width=7.0cm,height=5cm]{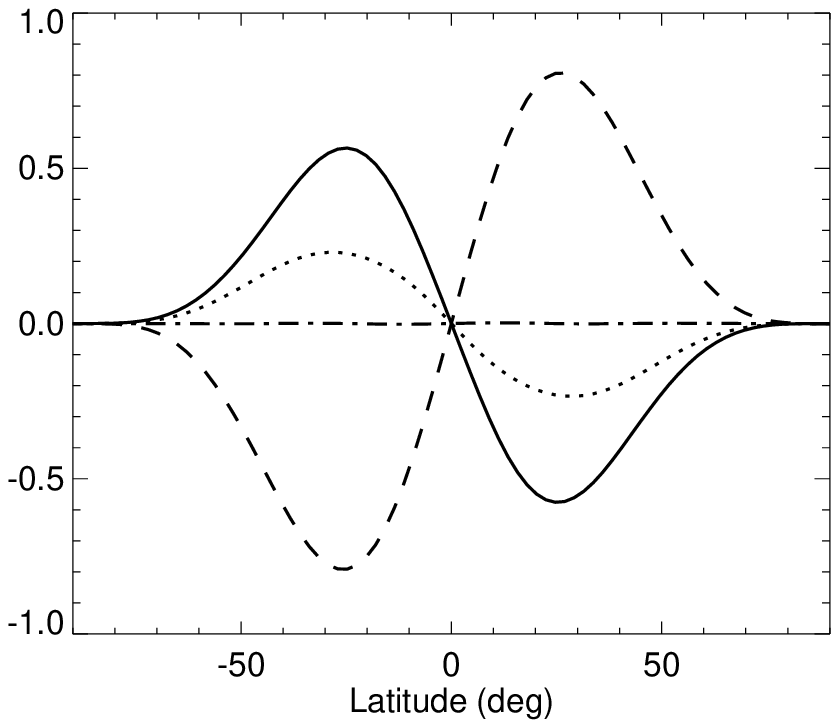} \\
  \includegraphics[width=7.0cm,height=5cm]{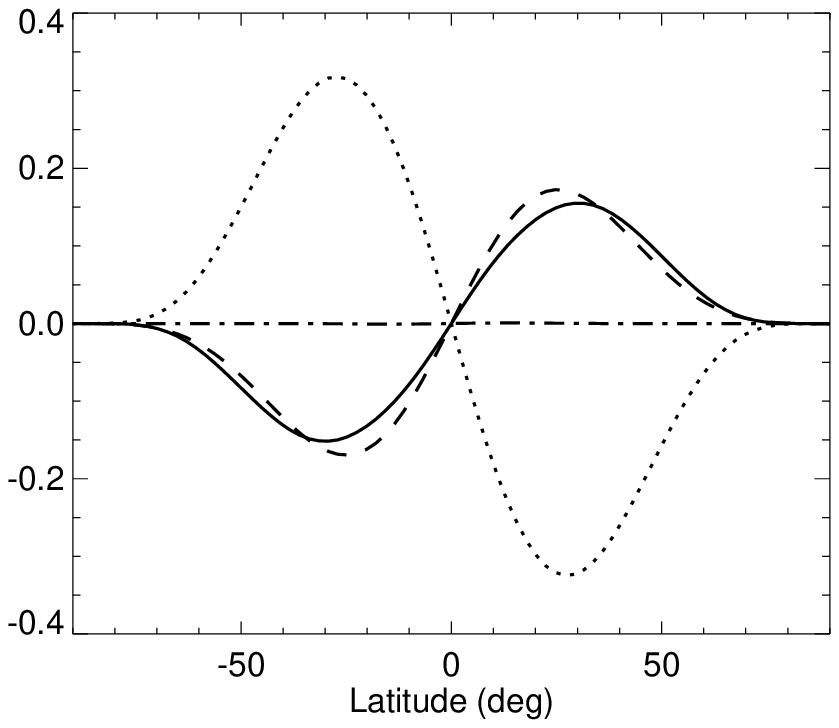}
  \includegraphics[width=7.0cm,height=5cm]{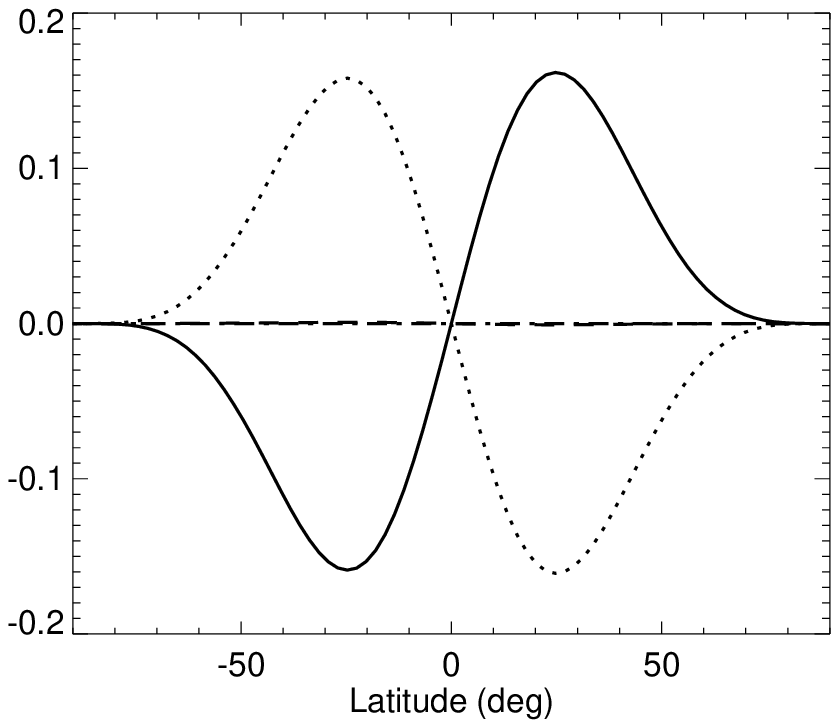}
 \end{center}
  \caption{ Horizontal flux of angular momentum for the F star with $c_\nu=0.15$ and rotation periods of 4 d (top left), 10 d (top right), 27 d (bottom left), and 100 d (bottom right). The line styles are the same as in Fig.~\ref{ftet_sun}.
    \label{ftet_fstar}
    }
\end{figure*}
The anisotropy of the convective heat transport plays a crucial part in maintaining the latitudinal shear. The baroclinic term that arises from the horizontal temperature gradient counteracts the shear along the $z$-axis  as a driver of meridional flow.

Brun and Toomre (2002) have compared the contributions of the \L-effect, viscosity, and meridional flow to the total horizontal flux of angular momentum,
\begin{equation}
  I(\theta) = \int^{\rm top}_{\rm bot} { F}_\theta(r,\theta) r \sin \theta dr,
\end{equation}
where ${ F}_\theta$ is the horizontal component of the local angular momentum flux vector.
Figure \ref{ftet_sun} shows the results from our solar model. The similarity between the  the full model and the corresponding diagrams in Fig.~11 of Brun \& Toomre (2002) is striking, especially for their model AB, where the total horizontal transport is smallest. The \L-effect is the most powerful transporter of angular momentum. It is balanced by the meridional flow and the viscous transport together, which are of comparable strength with the meridional flow being slightly stronger.

Figure \ref{ftet_fstar} shows how the balance of the three transport mechanisms
shifts as the rotation rate varies for the F8 star. For fast rotation (4 d) the
pattern is almost the same as for the solar-type star with the baroclinic term
switched off. The meridional flow and the \L-effect roughly cancel out each
other. The viscous part of the stress tensor is of minor importance. The result
for a period of 10 d looks very similar to the  Fig.~\ref{ftet_sun}. The
\L-effect dominates and the flow is stronger than the viscosity, though the
latter is no longer negligible. For the solar rotation rate of \mbox{27 d} the viscous transport has become dominant. The \L-effect and the meridional flow are of equal strength. Note that the contribution of the meridional flow has changed its sign. For very slow rotation (100 d) the horizontal \L-effect is negligible. The meridional flow (driven by the radial shear) is balanced by the viscous transport (Kippenhahn 1963).
%
%
\section{Conclusions}
We have applied the mean-field theory of angular momentum transport to the convection zone of a  main sequence star of spectral type F8. The only free parameter is the viscosity parameter $c_\nu$ which is calibrated by applying the model to the case of solar differential rotation. For rotation periods in the observed range the resulting rotation pattern  is similar to the solar one, though the horizontal shear is about twice as strong.

The shear vanishes in both the limiting cases of very fast and very slow rotation.
For fast rotation the Taylor-Proudman theorem requires  the rotation rate as constant on cylindrical surfaces. The only way to fulfill this constraint together  with the boundary conditions is rigid rotation.
For very slow rotation the horizontal \L-effect vanishes and the rotation rate is a function of the radius only.

The comparison of the contributions of the \L-effect, turbulence viscosity, and meridional flow shows that a horizontal differential rotation can only be maintained if the \L-effect is the dominant transporter. To maintain a stationary rotation pattern, it must be balanced by the contributions from the flow and the viscous transport. Since it requires shear to be effective while the flow does not, it is also necessary that viscosity plays a major part in establishing the balance. A meridional flow driven by the shear alone will always counteract the force generating the shear. Thus, the angular momentum transport by the flow will be directed against that by the \L-effect which generates the shear. For very large Taylor numbers the flow dominates over the viscous transport and the shear is strongly reduced. If, on the other hand, the flow is suppressed the angular momentum transport by the \L-effect has to be balanced by the viscosity, which requires differential rotation to be effective.

The solutions we find for the Sun and the F-star represent an intermediate
state. The contributions from the flow and the viscous part of the Reynolds
stress are of similar magnitude. Though not negligible, the meridional flow is
not strong enough to balance the \L-effect, leaving roughly half of the load to
the viscous transport. Consequently, the model of K\"uker \& Stix yields a value
of 0.2 only for $\delta \Omega/\Omega$, rather than the 0.3 of the simpler model
by K\"uker et al.~(1993), where the Reynolds stress is the only transporter of angular momentum. We have therefore readjusted the model by choosing $c_\nu=0.15$ instead of 4/15, thus requiring more shear to maintain the same flux of angular momentum.

The results suggest that the surface horizontal shear is mainly a function of the convection velocity, hence of the stellar luminosity. Our model does not produce a power law-type relation between rotation period and surface shear. Local fits yield exponents that are greater than those found by observers, but the scatter in the observations makes a direct comparison impossible.

%
%

%
\end{document}